\documentclass[12pt]{iopart}
\usepackage{graphicx}
\usepackage{dcolumn}
\usepackage{bm}
\usepackage{psfrag}
\usepackage{epsfig}
\usepackage{iopams}
\usepackage{amssymb}
\usepackage{color}
\usepackage{mathbbol}

\newcommand{\ehoch}{\ensuremath{\mathrm{e}^}}
\newcommand{\ehochi}[1]{\ensuremath{\mathrm{e}^{\mathrm{i}#1}}}
\newcommand{\ehochmi}[1]{\ensuremath{\mathrm{e}^{-\mathrm{i}#1}}}
\newcommand{\I}{\ensuremath{\mathrm{i}}}
\renewcommand{\Tr}{\ensuremath{\mathrm{Tr}}}

\newcommand{\bra}[1]{\ensuremath{\langle#1|}}
\newcommand{\ket}[1]{\ensuremath{|#1\rangle}}

\newcommand{\om}{\ensuremath{\omega_{\mathrm{m}}}}
\newcommand{\gmj}{\ensuremath{g_{\mathrm{m},j}}}

\newcommand{\gammam}{\ensuremath{\gamma_{\mathrm{m}}}}

\newcommand{\xzpm}{\ensuremath{x_{\mathrm{ZPM}}}}

\newcommand{\bplus}{\ensuremath{b^{\dag}}}

\renewcommand{\P}{\ensuremath{P}}

\def\pM{\ensuremath{\stackrel{+}{\scriptstyle(\kern-1pt-\kern-1pt)}}}
\def\mP{\ensuremath{\genfrac{}{}{0pt}{1}{-}{\scriptstyle(\kern-1pt+\kern-1pt)}}}

\graphicspath{{./graphics/}}

\begin{document}

\title{Steady-state negative Wigner functions of nonlinear nanomechanical oscillators}

\author{S.~Rips, M.~Kiffner, I.~Wilson-Rae and M.~J.~Hartmann}

\address{Technische Universit{\"a}t M{\"u}nchen, Physik Department,
James Franck Str., 85748 Garching, Germany\eads{\mailto{ignacio.wilson-rae@ph.tum.de}, \mailto{michael.hartmann@ph.tum.de}}}

\begin{abstract}
We propose a scheme to prepare nanomechanical oscillators in nonclassical
 steady states, characterized by a pronounced negative Wigner function. In our optomechanical approach, the mechanical oscillator couples to multiple laser driven resonances of an optical cavity. By lowering the resonance frequency of the oscillator via an inhomogeneous electrostatic field, we significantly enhance its intrinsic geometric nonlinearity per phonon. This causes the motional sidebands to split into separate spectral lines for each phonon number and transitions between individual phonon Fock states can be selectively addressed. We show that this enables the preparation of the nanomechanical oscillator in a single phonon Fock state. Our scheme can for example be implemented with a carbon nanotube dispersively coupled to the evanescent field of a state of the art whispering gallery mode microcavity.
\end{abstract}

\maketitle


\section{Introduction}
The manipulation of mechanical degrees of freedom by light scattering, presently known as optomechanics, has attracted
increasing interest in recent years \cite{Vahala,MarquardtGirvin09}. Current research in this direction is largely driven by
substantial progress in cooling mechanical oscillators to their ground states
\cite{Wilson-Rae2007,Marquardt2007,Rocheleau10,OConnell10,Teufel10,Chan11} and performing displacement measurements close to the standard
quantum limit \cite{Teufel09,Schliesser09,Anetsberger10}. These developments set the stage for investigations of the quantum regime of
macroscopic mechanical degrees of freedom, an endeavor that could lead to experimental tests of potential limitations of 
quantum mechanics \cite{Adler}. Here, we introduce an optoelectromechanical technique to prepare a mechanical oscillator in a nonclassical steady state with a negative Wigner function by enhancing the anharmonicity of its motion.

As has been known since the initial formulation of quantum mechanics, the dynamics of a purely harmonic quantum system
is hard to distinguish from its classical counterpart \cite{ehrenfest}. As an example, electrical circuits in the
superconducting regime do not display quantum behaviour unless they feature a nonlinear element such as a Josephson
junction. Along these lines nanomechanical oscillators that are coupled to nonlinear ancilla systems have been
considered \cite{ABS02,Jacobs,OConnell10}.
Here, in contrast, we introduce a framework that only makes use of intrinsic properties of the mechanical oscillator.
This approach allows one to prepare the latter in nonclassical states that are stationary states of its dissipative motion.
The lifetime of these states is thus not limited by the lifetime of the oscillator's excitations.
 
In fact, doubly clamped mechanical resonators such as nanobeams 
naturally feature
an intrinsic, geometric nonlinear contribution to their elastic energy
that gives rise to a Duffing nonlinearity \cite{Zeldovich1974,Dykman1978,Duffing06,Babourina} and typically only becomes relevant for
large deflection amplitudes. 
To investigate quantum effects of mechanical degrees of freedom \cite{Eisert08,Katz07,hartmann08} it is thus of
key importance to devise means for enhancing the nonlinearity of nanomechanical structures.

\begin{figure}
\includegraphics[width=0.9\textwidth]{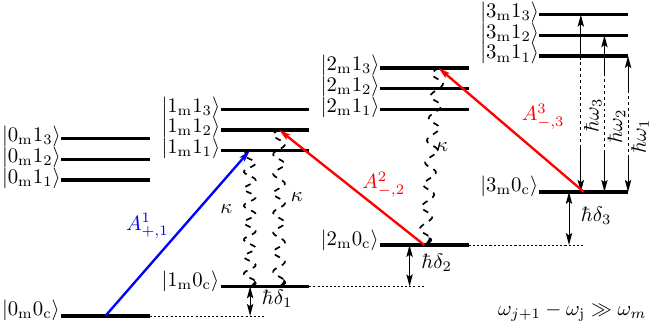}
\caption{Level diagram of the coupled
  cavity-resonator-system in a shifted representation.
  $\ket{n_{\mathrm{m}}1_{j}}$ denotes a single photon state
  of cavity mode $j$ with $n$ phonons in the mechanical mode,
  while $|0\rangle_c$ denotes the vacuum (i.e.~the steady state
  of the decoupled cavity). Due to the nonlinearity, the
  mechanical energy level spacings depend on $n$ and can be
  addressed individually. Only resonant processes are sketched
  with a blue detuned laser acting on the transition
  $\ket{0}_{\mathrm{m}}\rightarrow\ket{1}_{\mathrm{m}}$ and red
  detuned lasers acting on transitions
  $\ket{2}_{\mathrm{m}}\rightarrow\ket{1}_{\mathrm{m}}$ and
  $\ket{3}_{\mathrm{m}}\rightarrow\ket{2}_{\mathrm{m}}$.}
 \label{niveauschema}
\end{figure}

Here we consider inhomogenous electrostatic fields to decrease the harmonic oscillation frequency of a nanomechanical oscillator \cite{Unterreithmeier}. 
This procedure enhances the amplitude of its zero point motion and therefore leads to an amplification of its
nonlinearity per phonon, which eventually becomes comparable to the optical linewidth of a high finesse cavity. 
When combined with an optomechanical coupling to the cavity, this approach opens up a variety of possibilities for
manipulating the mechanical oscillator by driving the cavity modes with lasers. In particular the motional sidebands split
into separated lines for each phonon number $n$, c.f. figure \ref{niveauschema}. This feature allows to selectively transfer population from phonon Fock state $\ket{n}$ to $\ket{n+1}$ ($\ket{n'}$ to $\ket{n'-1}$) by driving the spectral line corresponding to $n$ ($n'$) phonons in the blue (red) sideband.

To demonstrate the capabilities of our approach, we show that suitably adjusted laser fields prepare the mechanical oscillator in a stationary state close to a phonon Fock state, i.e. in a state for which the quantum behaviour of an oscillator is most obvious \cite{hofheinz}.
The highly nonclassical nature of this state manifests itself in a Wigner function that reaches the lowest possible value of $-2/\pi$.
This can be verified via the power spectrum of an optical probe field \cite{Dykman1973,Dykman2011}. Here, the stationary nature of the motional states considerably simplifies their measurement as compared to phonon Fock states that might be prepared by projective measurements \cite{santamore,Jayich08,Buks08}. 

Our scheme can, for example, be implemented with state of the art nanoelectromechanical systems (NEMS) comprising carbon-based mechanical resonators \cite{Lassagne,Steele} 
dispersively coupled to high finesse
toroidal \cite{anetsberger09} or fiber based \cite{poellinger09} microcavities
(see figure \ref{setup}).
Carbon-based oscillators are particularly well-suited as they combine very small transverse 
dimensions with ultra-low dissipation \cite{Huettel09}.

The remainder of the paper is organized as follows. In section 2, we discuss the Hamiltonian of the oscillating nanobeam, including its geometric nonlinearity. In section 3, we then introduce the optomechanical coupling of the beam to a high finesse cavity and present the master equation that describes the dynamics of our system including the relevant damping mechanisms. In the following section 4, we provide an analysis of how driven cavity modes can drive the nanomechanical resonator into a non-classical steady state by adiabatically eliminating the photon degrees of freedom. In the sequel we analyse important aspects of an implementation in more detail. In section 5 we describe the electrostatic technique that lowers the harmonic oscillation frequency of the beam and hence increases its nonlinearity per phonon. The following section 6 describes a possible feasible experimental setup including a carbon nanotube coupled to a microtoroidal cavity. We then present examples of states that could be prepared in this setup in section 7. Finally, we discuss a technique to measure the prepared state (section 8) and close the paper with our conclusions and an outlook. 

\section{Oscillating nanobeam}
The dynamics of the fundamental flexural mode of a thin beam can be described by the Hamiltonian \cite{Phys.Rev.B64_220101},
\begin{equation} \label{rodmodel}
 H_\mathrm{m}=\frac{P^2}{2m_{*}}+\frac{1}{2}m_{*}\om^2X^2+\frac{\beta}{4}X^4\,,
\end{equation}
where $X$ is the deflection, $P$ its conjugate momentum, $\om$ the mode frequency and $m_{*}$ the effective mass. 
We assume clamped-clamped boundary conditions and apply thin beam theory \cite{Landau}. This yields
$\omega_{\mathrm{m},0}=c_{s}\tilde{\kappa}(\frac{4.73}{L})^2$ for the mode frequency without applying elastic stress or other external forces, and $\beta = 0.060 \, m_{*} \omega_{\mathrm{m},0}^{2}\tilde{\kappa}^{-2}$ for the nonlinearity which
results from the streching induced by the deflection.
Here, $L$ is the suspended length, $c_{s}$ is the phase velocity of compressional phonons and $\tilde{\kappa}$ is given by the relevant transverse dimension of the beam times a geometric prefactor.

The single-mode model (\ref{rodmodel}) is valid for sufficiently low phonon
occupancies and weak nonlinearities such that phonon-phonon interactions involving higher harmonics can be neglected.
Thus we focus on the latter regime and write $H_{\mathrm{m}}$ in terms of phonon creation (annihilation) operators $\bplus$ ($b$) to obtain
\begin{equation}\label{rodmodel_phonons}
H_{\mathrm{m}} = \hbar \om\bplus b+\hbar\frac{\lambda}{2}\left(\bplus+b\right)^4
\end{equation}
with nonlinearity $\lambda= \frac{\beta}{2} x_{\mathrm{ZPM}}^{4}/\hbar$ where
$x_{\mathrm{ZPM}}=\sqrt{\hbar/2m_{*}\om}$ is the amplitude of the oscillator's zero point motion.

Note that $\lambda \propto \om^{-2}$ can be significantly enhanced by ``softening'' the mode, i.e.~lowering $\om$.
Whereas this can be achieved by applying compressive stress along the beam \cite{Phys.Rev.B64_220101,Europhys.Lett.65_158-164},
we follow here an alternative approach to allow for better tunability.
We consider an in-plane flexural resonance (cf.~figure~\ref{setup}b) and use tip electrodes to apply a strong inhomogeneous static electric field which polarizes the beam.
As a dielectric body is
attracted towards stronger fields, this adds an inverted parabola to the potential that counteracts the harmonic
contribution due to elasticity. Whence, as will be borne out in detail in section 5, the oscillation frequency is reduced \cite{wu11,Unterreithmeier,Lee10}. 

\section{Optomechanics} We consider a typical optomechanical setup where the displacement
$X = \xzpm(b + \bplus)$ of the nanomechanical oscillator dispersively couples to several laser driven cavity resonances. In a frame rotating with the frequencies of the laser fields $\omega_{\mathrm{L},j}$ the system Hamiltonian reads
\begin{equation}
 H=\hbar \sum_{j}\left[-\Delta_ja_j^{\dag}a_j+\frac{\Omega_j}{2}\left(a_j^{\dag}+a_j\right)+G_{0,j}\xzpm a_j^{\dag}a_j\left(\bplus+b\right)\right]+H_{\mathrm{m}}\,,
\end{equation}
where $\Delta_j=\omega_{\mathrm{L},j}-\omega_j$ is the detuning
of a given laser $j$ from the closest cavity mode with frequency
 $\omega_j$ and photon annihilation operator $a_{j}$, $\Omega_j / 2 = \sqrt{P_{\mathrm{in},j}\kappa_{\mathrm{ex}}/\hbar\omega_{\mathrm{L},j}}$ is the driving strength associated to that laser with input power $P_{\mathrm{in},j}$ and $\kappa_{\mathrm{ex}}$ is the corresponding external decay rate. The coupling strength $G_{0,j}=\frac{\partial \omega_{j}}{\partial X}$ corresponds to the shift of the respective cavity resonance
per oscillator's deflection \cite{wilson-rae2008}.

We incorporate the damping of the cavity field at rate $\kappa$ and mechanical dissipation at rate $\gammam$ 
in a master equation and shift the cavity normal coordinates to their steady state value $a_j\rightarrow a_j+\alpha_j$, with
\begin{equation}
 \alpha_j=\frac{\Omega_j}{2 \Delta_j + \I \kappa}\,.
\end{equation}
Neglecting higher order terms in the photon-phonon coupling for the regime where $\langle a_{j}^{\dag} a_{j} \rangle \ll
|\alpha_{j}|^{2}$ \cite{Wilson-Rae2007,Marquardt2007,wilson-rae2008} leads to the linearized Hamiltonian 
\begin{equation}
\label{Hamiltonian_lin}
H'=\hbar\sum_j\left[-\Delta_ja_j^{\dag}a_j+\left(\frac{\gmj^*}{2}a_j+\mathrm{H.c.}\right)\left(\bplus+b\right)\right]+H_{\mathrm{m}}\nonumber\,,
\end{equation}
where $\gmj=2\alpha_j\xzpm G_{0,j}$ is the enhanced optomechanical coupling.
The term $\sim|\alpha|^2(\bplus+b)$, causing a static shift of the mechanical mode due to radiation pressure, is omitted here since it can be compensated by a suitable choice of electrostatic softening fields (see section 5). The master equation for the shifted system now reads
\begin{equation}
   \dot\rho=-\frac{\I}{\hbar}\left[H',\rho\right]+\sum_j\frac{\kappa}{2}\mathcal{D}_{\mathrm{c},j}\rho+\frac{\gammam}{2}\mathcal{D}_{\mathrm{m}}\rho\,,\\
\label{mg}
\end{equation}
where 
\begin{eqnarray}
\label{damping1}\mathcal{D}_{\mathrm{c},j}\rho=&2a_j\rho{a_j}^{\dag}-{a_j}^{\dag}a_j\rho-\rho{a_j}^{\dag}a_j\,,\\
\nonumber\mathcal{D}_{\mathrm{m}}\rho=&(\overline n+1)\left\{2b\rho b^{\dag}-b^{\dag}b\rho-\rho b^{\dag}b\right\}\\
\label{damping2}&+\overline n\left\{2b^{\dag}\rho b- bb^{\dag}\rho-\rho bb^ {\dag}\right\}\,,
\end{eqnarray}
with the Bose number $\overline n=1/\left[\exp(\hbar\om/k_{\mathrm{B}}T)-1\right]$ at environment temperature $T$.
In equation (\ref{damping2}) we have applied a rotating wave approximation that is justified for the high mechanical quality factors we are interested in \cite{wilson-rae2008}.

Equation (\ref{mg}) describes the dynamics of our system. Yet, the behaviour of the mechanical mode becomes more apparent after adiabatically eliminating the photon degrees of freedom, as we do in the next section.

\section{Reduced master equation}
For $|\gmj|\ll\kappa$ we now derive a reduced master equation (RME) for the mechanical motion by adiabatically eliminating
the cavity degrees of freedom \cite{wilson-rae2008}. We start by applying a rotating wave approximation (RWA) to the mechanical Hamiltonian
which yields, 
\begin{equation}\label{rodmodel_phonons_rwa}
 H_{\mathrm{m}}'= \hbar\om'\bplus b+\hbar\frac{\lambda'}{2}\bplus\bplus bb\,,
\end{equation}
with the shifted frequency $\om'=\om+\lambda'$ and the nonlinearity per phonon $\lambda'=6\lambda$. Here, the RWA is warranted provided the relevant phonon numbers $n$ fulfil $n^2\lambda' \ll 6\om$. 
Within the RWA, the mechanical energy eigenstates are Fock states and the large nonlinearity,
$\lambda' \gg |\gmj|^{2}/\kappa, \overline{n} \gammam$, we are interested in naturally leads to Fock number resolved dynamics.
In an interaction picture with respect to $H_{\mathrm{m}}'$ transition operators between neighbouring phonon Fock states rotate at frequencies corresponding to the respective mechanical energy level spacings, as
\begin{equation}
\ehochi{H_{\mathrm{m}}' t/\hbar} b \ehochmi{H_{\mathrm{m}}' t/\hbar} = \sum_n\ehochmi{\delta_{n}t}b_n\,,
\end{equation}
with $b_n=\sqrt{n}\ket{n-1}\bra{n}$ and 
\begin{equation}
 \delta_n=(E_n-E_{n-1})/\hbar\approx\om'+\lambda'(n-1)\,.
\end{equation}
We further focus on the resolved sideband regime for the softened frequency $\kappa\ll\om$, which is suitable for efficient back action cooling, as well as $\overline{n}\gammam\ll|\gmj|^2/\kappa$ to ensure that the influence of the cavity modes on the mechanical motion overwhelms environmental heating. The resulting RME for the reduced density operator of the nanomechanical resonator,
$\mu=\Tr_{\mathrm{c}}\left\{\rho\right\}$, reads
\begin{equation}
\label{reduced mg}
\dot\mu=-\frac{\mathrm{i}}{\hbar}\left[H_{\mathrm{m}}'+\hbar\sum_{n,j}\Delta_{\mathrm{m},j}^{(n)}b_n^{\dag}b_n,\mu\right] +\sum_{n,j} \sum_{\sigma=+,-} \frac{A_{\sigma,j}^n}{2} \, \mathcal{D}_{\sigma}^n\mu + \frac{\gammam}{2}\mathcal{D}_{\mathrm{m}}\mu\,,
\end{equation}
where the driven cavity modes act like an additional bath for each number transition with damping terms 
\begin{eqnarray}
 \mathcal{D}_+^n\mu&=2b_n^{\dag}\mu b_n-b_nb_n^{\dag}\mu-\mu b_nb_n^{\dag}\,,\\
\mathcal{D}_-^n\mu&=2b_n\mu b_n^{\dag}-b_n^{\dag}b_n\mu-\mu b_n^{\dag} b_n\,.
\end{eqnarray}
Note that the frequency shifts 
\begin{equation}
\Delta_{\mathrm{m},j}^{(n)}=\gmj^2\left\{\frac{\Delta_j+\delta_n}{4(\Delta_j+\delta_n)^2+\kappa_j^2}+\frac{\Delta_j-\delta_n}{4(\Delta_j-\delta_n)^2+\kappa_j^2}\right\}
\end{equation}
are negligible compared to $\om',\lambda'$.
The cavity-induced transition rates between neigbouring Fock
states 
\begin{equation}
 A_{\pm,j}^n=\frac{|\gmj|^2 \kappa}{4(\Delta_j\mp\delta_n)^2+\kappa^2}
\end{equation}
however, can exceed the mechanical damping $A_{\pm,j}^n>\gammam \overline n$ which leads to a significant impact on the mechanical steady state.

The steady state of the RME (\ref{reduced mg}) is diagonal in
the Fock basis,
\begin{equation}
\mu=\sum_nP_n\ket{n}\bra{n}\,.
\end{equation}
Hence the relations 
\begin{equation}
 \frac{P_n}{P_{n-1}}=\frac{\sum_jA_{+,j}^n+\gammam\overline{n}}{\sum_jA_{-,j}^n+\gammam\left[\overline{n}+1\right]}\,,\:\hspace{2cm}\:\sum_nP_n=1\,,
 \label{steadystatecondition}
\end{equation}
determine all occupation probabilities $P_n$ and therefore the mechanical state $\mu$. The relations (\ref{steadystatecondition}) show the central feature
of the driven nonlinear nanoresonator.
Here $|\gmj|^2/\kappa \gg \gammam\overline{n}$ implies
$P_n/P_{n-1} \approx (\sum_jA_{+,j}^n)/(\sum_jA_{-,j}^n)$. For a
detuning $\Delta_{j} = \pm\delta_n$ we find $A_{\pm,j}^n >
A_{\mp,j}^n$ and the corresponding laser drive therefore enhances (decreases) $P_n/P_{n-1}$.

This feature can be used to approach phonon Fock states $\ket{n}$,
as illustrated in figure~\ref{niveauschema} for $n=1$.
Fock states are highly nonclassical and for $n$ odd they exhibit a maximally negative Wigner function at the origin.
Ensuring that each laser drive preferentially addresses only one transition $\ket{n}\rightarrow\ket{n\pm1}$ naturally requires
$\lambda' \gtrsim \kappa$. This can be achieved by electrostatic mode softening as we describe in the next section.

\section{Dielectric softening}

The harmonic frequency $\om$ of the nanoresonator can be reduced by adding an electrostatic potential $V_{\mathrm{es}}\propto-X^2$ to the elastic potential, where $X$ is the deflection of the resonator. This is achieved by placing charged electrodes at both sides of the nanoresonator (see figure \ref{setup}), that generate a strongly inhomogenous electrostatic field. To estimate the potential that softens the beam, we consider its electrostatic energy line density
\begin{equation}
 \mathcal{W}(x,y)=-\frac{1}{2}[\alpha_{\parallel}E^2_{\parallel}(x,y)+\alpha_{\perp}E^2_{\perp}(x,y)]\,.
\end{equation}
Here, $y$ and $x$ are the co-ordinates along the beam and the direction of its deflection, $E_{||,\bot}$ are the electrostatic field components parallel and perpendicular to the beam and $\alpha_{||,\bot}$ the respective screened
polarizabilities. The electrostatic energy as a functional of the deflection $x(y)$ thus reads 
\begin{equation}
V_{\mathrm{es}}=\int_0^L\mathcal{W}(x(y),y){\rm d}y\,.
\end{equation}
We expand $\mathcal{W}(x(y),y)$ to second order in the transverse deflection $x(y)$ from its equilibrium configuration $x(y)=0$.
The leading contributions for the fundamental mode with profile $\phi_{0}(y)$ can be found by considering
$x(y)=\phi_{0}(y) X$. They read
\begin{equation}
V_{\mathrm{es}} \approx V_{\mathrm{es},0} + V_{\mathrm{es},1} X + V_{\mathrm{es},2} X^{2}\,,
\end{equation}
where $V_{\mathrm{es},0}$ is an irrelevant constant and 
\begin{eqnarray}
V_{\mathrm{es},1}&=\int_{0}^{L}{\rm d}y\frac{\partial\mathcal{W}(x,y)}{\partial x}\big{|}_{x=0}\phi_0(y)\,,\\
V_{\mathrm{es},2}&=\frac{1}{2}\int_{0}^{L}{\rm d}y\frac{\partial^{2}\mathcal{W}(x,y)}{\partial x^{2}} \big|_{x=0}\phi^{2}_0(y)\,.
\end{eqnarray}
The linear term $V_{\mathrm{es},1} X$ shifts the beam's equilibrium position and can be employed to compensate shifts induced by the dispersive coupling to the cavity fields by satisfying $V_{\mathrm{es},1}=-\hbar\sum_jG_0|\alpha_j|^2$. The quadratic term $V_{\mathrm{es},2}
X^{2}$ in turn is negative and reduces the fundamental frequency as 
\begin{equation}
\om^{2} = \omega_{\mathrm{m},0}^{2} - \frac{2}{m_{*}}
|V_{\mathrm{es},2}|\,.
\end{equation}
 At $|V_{\mathrm{es},2}| = \frac{m_{*}}{2}\omega_{\mathrm{m},0}^{2}$ the buckling instability \cite{Phys.Rev.B64_220101,Europhys.Lett.65_158-164} occurs and by tuning the electrostatic fields below it one can reduce $\om$ by
any desired factor $\zeta = \omega_{\mathrm{m},0}/\om$. Note that the softening increases the zero point motion amplitude and, hence, also the
optomechanical coupling per photon $G_0\xzpm$.  A particularly useful feature of this dielectric approach to soften the
nanobeam is that the electrostatic potential and hence $\om$, $G_0\xzpm$ and $\lambda$ can be tuned in situ. To illustrate the feasibility of our approach we now turn the description of a feasible experimental setup for its implementation.

\section{Carbon nanotube and microtoroid}
\begin{figure}
 \centering
\includegraphics[width=0.9\textwidth]{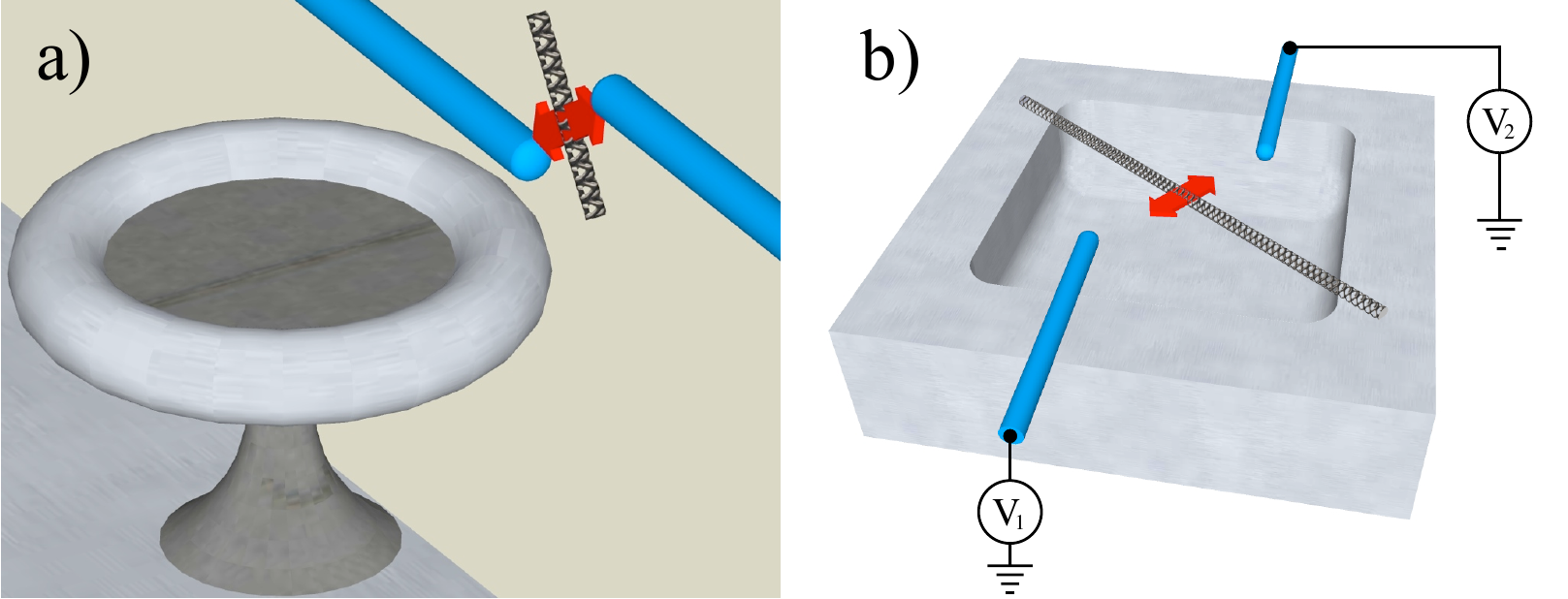}
\caption{ {\bf a)}
  Possible setup with a carbon nanotube dispersively coupled to
  the evanescent field of a microtoroid and tip electrodes used
  for the electrostatic softening of the mechanical frequency
  (blue cylinders). {\bf b)} NEMS chip carrying the nanotube and
  the electrodes. The oscillation direction of the nanotube is
  indicated by the red arrows. In {\bf a)}, nanotube and
  electrodes are not drawn to scale with the toroid and the NEMS
  chip is not shown for illustration purposes. The placement of nanotube and electrodes
  minimizes degradation of the cavity finesse and
  maximizes optomechanical coupling.}
\label{setup}
\end{figure}

As a suitable device for realizing our scheme we envisage a setup with a single-walled carbon nanotube as the
mechanical oscillator ($c_{s}=21000 \mathrm{m}\mathrm{s}^{-1}$) that interacts with the evanescent field of a whispering gallery mode
of a microtoroid cavity \cite{anetsberger09}, cf.~figure.~\ref{setup}.  A carbon nanotube is a favorable
candidate since without softening ($\om = \omega_{\mathrm{m},0}$) one obtains $\lambda\propto (m_{*} \tilde{\kappa}^2)^{-1}$
indicating that low effective mass $m_{*}$ and small transverse dimension $\propto \tilde{\kappa}$ are desirable. For a carbon nanotube of radius $R$, 
one has $\tilde{\kappa}=R/\sqrt{2}$.
We consider a wide band gap semiconducting carbon nanotube which for the infrared laser-wavelengths involved will behave as a dielectric
\cite{Wilson-Rae09}.  Taking into account that the transverse optical polarizability is negligible (i.e.~depolarization) and
approximating the longitudinal one by its static value,
the optomechanical coupling can be estimated as (cf.~\cite{anetsberger09})
\begin{equation}
G_{0,j}\sim\omega_j \frac{\alpha_{\parallel}
  L}{\epsilon_0V_{\mathrm{c}}}
\xi^2\kappa_{\perp}\ehoch{-2\kappa_{\perp}d}\, C,
\end{equation}
where $V_{\mathrm{c}} = \pi a_{\mathrm{c}}^{2} L_{\mathrm{c}}$ is the mode
volume of the whispering gallery mode with cavity length $L_{\mathrm{c}}$ and
effective mode waist $2 a_{\mathrm{c}}$. $\xi$ is the ratio of the field
at the cavity surface to the maximum field inside, $d$ the
distance between the chip carrying the nanotube and the microtoroid's surface, and
$\kappa_{\perp}^{-1}$
the decay length of the evanescent field.
The correction factor $C$ depends on the type
of whispering gallery mode and the placement of the carbon nanotube. We assume for
simplicity that the NEMS and photonic chips are perpendicular to
each other and find for a TE mode
and optimal placement $C\approx0.17/\sqrt{\kappa_{\perp}(d+a_c)}$.
This optimum involves a small displacement of the
carbon nanotube midpoint from the equatorial plane of the toroid to
ensure coupling to the in-plane motion and an angle of $\sim
50^{\circ}$ between the carbon nanotube axis and the equatorial plane of the toroid to circumvent
depolarization.

In this setup, the tip electrodes used to lower the mechanical
oscillation frequency need to be positioned close to the cavity
surface to ensure a sufficiently inhomogenous electrostatic
field while maintaining the optomechanical coupling high enough.
Hence, a concomitant degradation of the cavity's finesse
due to photon scattering or absorption at the tip
electrodes is a natural concern.
We find that cavity losses are
minimized if one uses TE whispering gallery modes and tip electrodes of
sub-wavelength diameter $d_{\mathrm{el}}$ that are aligned
approximately parallel to the equatorial plane of the toroid (cf.~figure.~\ref{setup}a and
\ref{setup}b). 
For this arrangement the dominant loss
mechanism is absorption. The ratio of absorbed power $P_{a}$ to
power circulating in the cavity $P_{c}$ can be estimated to be (c.f. \cite{jackson}, chapter 8),
\begin{equation}
\frac{P_{a}}{P_{c}} \lesssim \frac{\pi \sigma d_{\mathrm{el}}
  \xi^{2}}{c \varepsilon_{0} a_{\mathrm{c}}}
\sqrt{\frac{\pi}{\kappa_{\perp} a_{\mathrm{c}}}} e^{-2 k_{\perp}
  d} \sin \theta\,,
\end{equation}
where $\sigma$ is the 2D optical conductivity
of the electrodes and $\theta$ their misalignment with respect
to the equatorial plane of the toroid. The resulting finesse is given by 
\begin{equation}
 \mathcal{F}
= \left(\frac{1}{\mathcal{F}_{c}} + \frac{P_{a}}{2 P_{c}}\right)^{-1}\,, 
\end{equation}
where $\mathcal{F}_{c}$ is the bare cavity finesse. For relevant cavity
parameters we find $\frac{P_{a}}{2 P_{c}}
\ll \mathcal{F}_{c}^{-1}$ if we assume $\theta \le 1^{\circ}$,
$d_{\mathrm{el}} \le 10\,$nm and $\sigma \lesssim 2 \times
10^{-5}\,\Omega^{-1}$. Metallic carbon nanotubes, for example, fulfil the
latter two requirements provided their resonances do not match
the relevant cavity frequencies $\omega_j$ \cite{Joh11}.

In the following we now give some examples of steady states for the mechanical oscillator that can be generated
in the setup described here.

\section{Examples}
\begin{figure}
\includegraphics[width=0.9\textwidth]{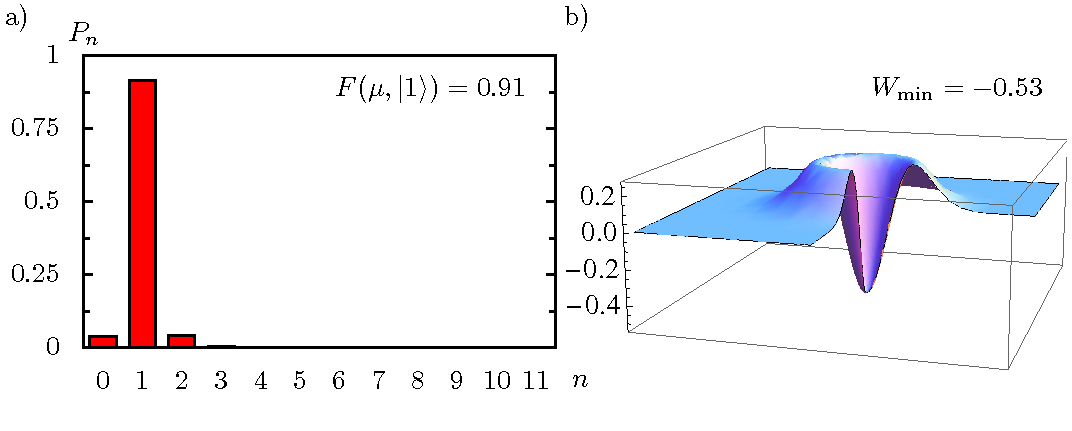}
\caption{Steady state of an oscillating (10,0) carbon nanotube with $L=1.0\,\mu$m and $Q_{\mathrm{m}}=5\times10^6$ for $T = 20\,$mK, $\mathcal{F}_{c}= 3 \times 10^6$ and $L_{\mathrm{c}}=1.35\,$mm. a) Fock state occupation probabilities $P_{n}$. b) Wigner function $W$.
These results are obtained from a numerical solution of equation (\ref{mg}).
}
\label{example1}
\end{figure}

To illustrate the potential of our scheme, we show in
figure~\ref{example1} results for a $(10,0)$ carbon nanotube 
characterised by $R=0.39\,$nm, $\alpha_\parallel= 142 (4\pi\epsilon_0$\AA$^2)$ and $\alpha_\perp= 10.9
(4\pi\epsilon_0$\AA$^2)$ \cite{Kozinsky}. The nanotube has a length of $L=1.0\,\mu$m
and is coupled to whispering gallery modes with wavelengths $2\pi
c/\omega_j\approx 1.1\,\mu$m, $a_{\mathrm{c}} = 1.4\,\mu$m, and
$\xi = 0.4$ of a silica microtoroid with finesse
$\mathcal{F}_{c}= 3 \times 10^6$ \cite{kippenberg2004} and
circumference $L_{\mathrm{c}}=1.35\,$mm. 
The applied voltages are chosen so
that tip electrodes at a distance of $20\,$nm from the tube
induce maximal fields at its axis of $E_{\parallel}\approx 1.20
\times 10^{7}\,\mathrm{V/m}$ and $E_{\perp}\approx 1.778\times 10^6\,$V/m
and lower its mechanical frequency by a factor $\zeta=4.0$.
For the optomechanical couplings, we consider
$g_{\mathrm{m},j}/2\pi=21.0\,$kHz for all $j$ which can be
achieved with $d = 50\,$nm, $\kappa_{\mathrm{ex}}/\kappa=0.1$ and
a launched power $P_{\mathrm{in},j}=1.2\,$W.
The above parameters lead to $\om/2\pi= 5.23\,$MHz,
$\lambda/2\pi= 209\,$kHz
and $\kappa/2\pi=52.3\,$kHz.
We assume carbon nanotube $Q$-values of $Q_{\mathrm{m}}=5\times10^6$ and $Q_{\mathrm{m}}=5\times10^5$
\footnote{For carbon nanotubes $\frac{\om}{2\pi} Q_{\mathrm{m}} \!
  \sim \! 10^{14}\,$Hz has already been achieved (G.A. Steele, private communication) and there are
  strong indications that mechanical dissipation in doubly clamped CNTs strongly depends on the amplitude of the motion \cite{eichler}. Hence even larger $Q$-values would result for the amplitudes comparable to the zero point motion that are relevant here.}
and an
environmental temperature $T = 20\,$mK.  

For these parameters and for laser detunings $\Delta_1 \approx \delta_1$ and $\Delta_{2,3} \approx -\delta_{2,3}$ a
single phonon Fock state can be prepared with 91\% fidelity (for $Q_{\mathrm{m}}=5\times10^6$) as shown in figure~\ref{example1}.
This steady state of the mechanical oscillator has a Wigner function that is clearly negative at the origin, $W(0,0) = -0.53$ for $Q_{\mathrm{m}}=5\times10^6$ and $W(0,0) = -0.30$ for $Q_{\mathrm{m}}=5\times10^5$. 

The results shown in figure~\ref{example1} have been obtained from a numerical solution of equation (\ref{mg}) with numerically optimised detunings $\Delta_j$. These numerical results do not rely on the rotating wave approximation used in equation (\ref{rodmodel_phonons_rwa}).
Here the fastly rotating terms lead to corrections to the energy level spacings $\delta_n$ that can be comparable to $\kappa$.
Hence to match the laser detunings with sufficient accuracy to the energy level spacings $\delta_n$, the latter have been determined numerically. 

\begin{figure}
\includegraphics[width=0.9\textwidth]{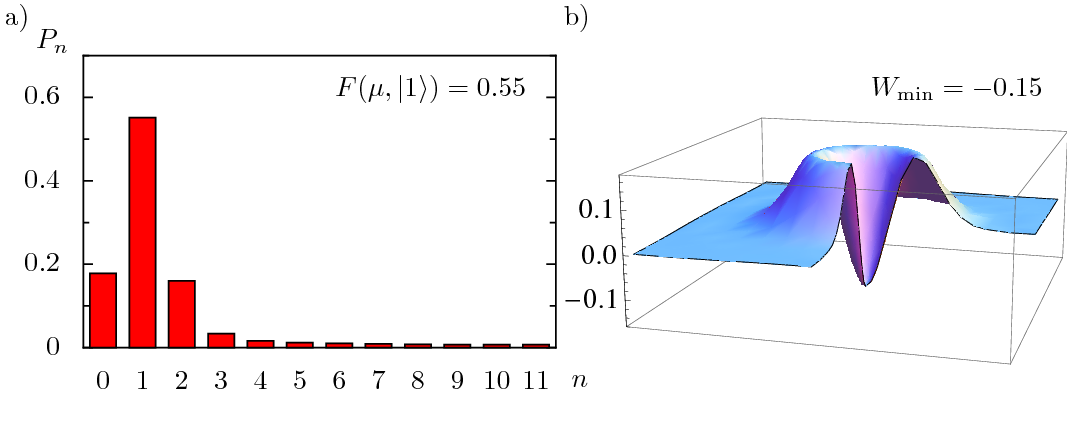}
\caption{Steady state of an oscillating (10,0) carbon nanotube with  $L=1.7\,\mu$m and $Q_{\mathrm{m}}=1.5\times10^6$ for $T = 30\,$mK, $\mathcal{F}_{c}= 2 \times 10^6$ and $L_{\mathrm{c}}=1.8\,$mm. a) Fock state occupation probabilities $P_{n}$. b) Wigner function $W$.
These results are obtained from a numerical solution of equation (\ref{mg}).}
\label{example2}
\end{figure}

In turn, for the more
moderate parameters: $L=1.7\,\mu$m,
$\mathcal{F}_{c}=2\times 10^6$,
$L_c=1.80\,$mm ($\kappa/2\pi=64.2\,$kHz),
$\zeta=3.3$, (which imply $\om/2\pi= 2.13\,$MHz,
and $\lambda/2\pi=85.5\,$kHz),
launched laser powers
$P_{\mathrm{in},1/2}=22\,$mW
and $P_{\mathrm{in},3}=44\,$mW,
$Q_{\mathrm{m}}=1.5\times10^6$,
and $T=30\,$mK,
we still find a significant negative peak of depth
$W_{\mathrm{min}}=-0.15$ (see figure~\ref{example2}). 

We note that the deep resolved-sideband and undercoupled conditions used here, where the frequencies of the driving lasers are detuned by several cavity linewidths from the optical resonances of the cavity and losses into the coupling fiber are only a small fraction of the total losses, imply that the power absorbed inside the cavity is much smaller than the launched power (below $30\,\mu$W in the example of figure \ref{example1} and below $7\,\mu$W in the example of figure \ref{example2}). Whereas an intracavity absorption $\sim1\,\mu$W has already been shown to be compatible with a cryogenic environment \cite{riviere11}, even higher finesses and lower absorption could be attained by using crystalline resonators \cite{Hofer10}. Additionally, the CNT polarizabilities could be enhanced well above the static values we have assumed by profiting from the corresponding excitonic optical resonances \cite{Htoon04} allowing for a substantial reduction in the laser powers required.

Furthermore, in both examples, the voltages applied to soften the frequency of the CNT nanomechanical resonator generate electric fields at the CNT that are well below its threshold for dielectric breakdown and are comparable to fields applied in recent experiments \cite{Unterreithmeier}.

Finally, we note that our approach can be generalized to the preparation of Fock states with $n>1$.

\section{Measurements}
To verify whether a nonclassical steady state with a negative Wigner function $W$ has been successfully prepared, one
can measure sidebands of the output power spectrum $S(\omega)$ of an additional cavity mode weakly driven by a probe laser with
frequency $\omega_{\mathrm{L}}$. Within our quantum noise approach \cite{wilson-rae2008} and under the same conditions for which
Eq. (\ref{reduced mg}) holds we find for the sidebands (i.e. where $\omega\neq\omega_L$)
\begin{equation}
\label{Spectrum}
S(\omega)\propto\sum_n \left[\frac{n \, \Gamma_n \, A_-^n \, P_{n}}{\left(\omega-\omega_{\mathrm{L}}-\delta_n\right)^2+\frac{\Gamma_n^{2}}{4}}
+ \frac{n \, \Gamma_n \, A_+^n \, P_{n-1}}{\left(\omega-\omega_{\mathrm{L}}+\delta_n\right)^2+\frac{\Gamma_n^{2}}{4}}\right]\,,
\end{equation}
with sideband linewidths
\begin{eqnarray}
 \nonumber\Gamma_{n}=&n [ A_-^n + A_+^n + \gammam (2\overline{n}+1)] \\
\nonumber &+(n-1)[A_-^{n-1} + \gammam (\overline{n}+1)]\\
&+(n+1) [A_+^{n+1}+\gammam\overline{n}]\,.
\end{eqnarray}
\begin{figure}
\includegraphics[width=0.9\textwidth]{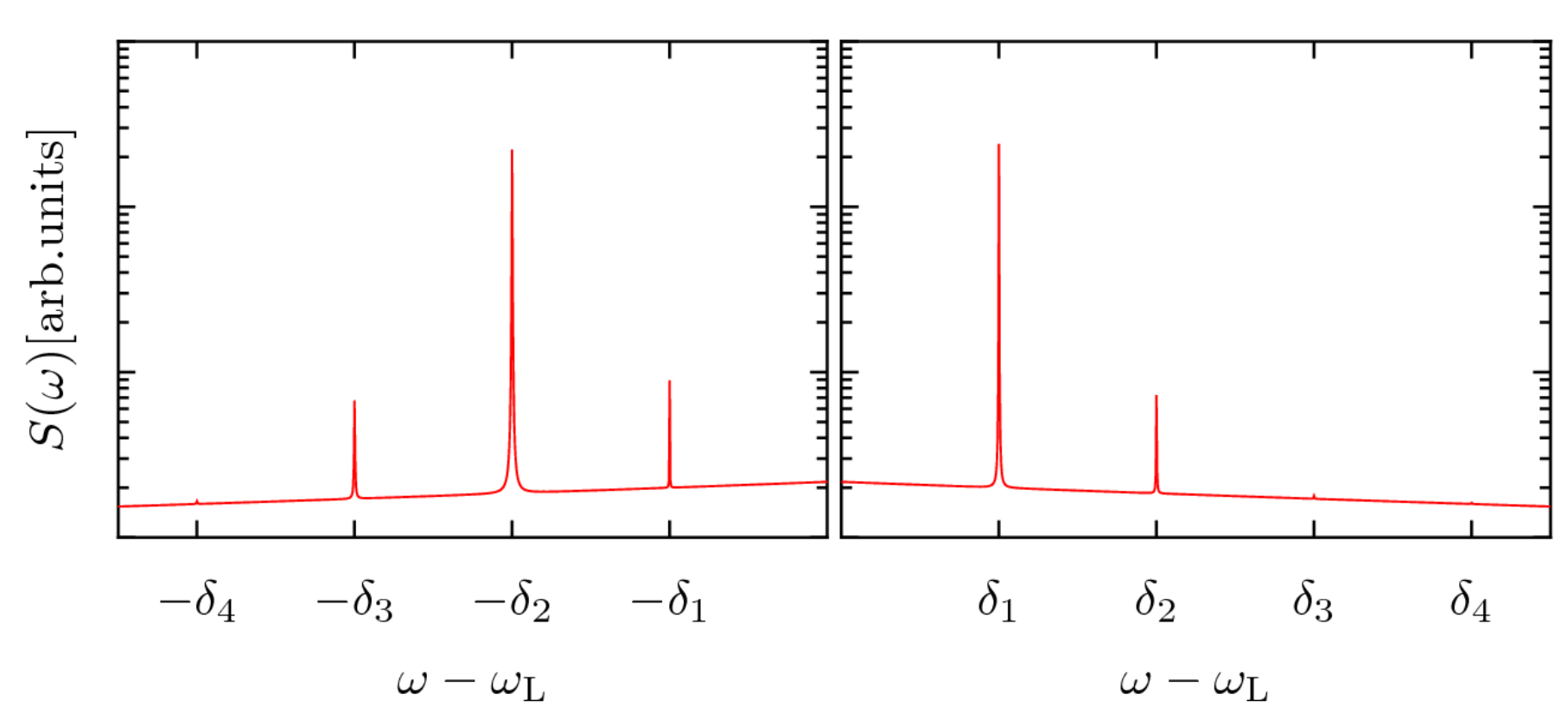}
\caption{Power spectrum $S(\omega)$ of a resonant probe laser with the sidebands split into well resolved lines corresponding to the phonon numbers $n$. This example shows the result of equation (\ref{Spectrum}) for the parameters also used in figure \ref{example1}.}
\label{spectrum_fig}
\end{figure}
Since the negativity of the Wigner function is in our scheme expected to be maximal at the origin where $W(0,0) = \frac{2}{\pi} \sum_{n} (-1)^{n} P_{n}$, the nonclassicality of the state can be read off from $S(\omega)$.
For a probe laser that is resonant with a cavity mode, one has $A_+^n=A_-^n$ and the ratios of peak intensities
$S(\omega_{\mathrm{L}}+\delta_n)/S(\omega_{\mathrm{L}}-\delta_n) \approx P_n/ P_{n-1}$ 
together with $1 = \sum_{n} P_{n}$
allow to determine the $P_{n}$ using that occupations above a certain Fock number are negligible.

\section{Conclusions and Outlook}
We have shown that a dielectrically enhanced nonlinearity in
nanomechanical oscillators enables controlling their motion at a
Fock state resolved level. The scheme set forth opens new
avenues for optomechanical manipulations of
nanobeams, in particular by exploiting the in situ tunability of
the electrostatic softening approach introduced.

\section{Acknowledgements}
This work is part of the DFG - Emmy Noether project HA 5593/1-1
and was supported by Nanosystems Initiative Munich (NIM). The
authors thank J.~Kotthaus, T.J.~Kippenberg, and A.~Schliesser
for enlightening discussions.

\end{document}